\documentstyle[preprint,aps]{revtex}
  
\begin{document}

\title{ 
	Low-frequency Raman scattering in model disordered solids: \\
	percolators above threshold.
	}
\author{ 
	O.~Pilla and G.~Viliani
	}
\address{ 
	  Dipartimento di Fisica and Istituto Nazionale di Fisica della 
	  Materia, \\ Universita' di Trento, I-38050 Povo, Trento, Italy
	  }
\author{
	 R.~Dell'Anna and G.~Ruocco
	 }
\address{ 
	  Dipartimento di Fisica and Istituto Nazionale di Fisica della  
	  Materia, \\ Universita' dell'Aquila, I-67100 Coppito, l'Aquila,
Italy
	  }
\maketitle
\date{\today}
\begin{abstract}
The Raman coupling coefficients of site- and bond-percolators at
concentration higher than percolation threshold are computed for
two scattering mechanisms: Bond Polarizability (BPOL) and 
Dipole-Induced-Dipole (DID). The results show that DID does not
follow a scaling law at low frequency, while in the case of
BPOL the situation is less clear. The numerically computed 
frequency dependence in the case of BPOL, which  can be
considered a good scattering mechanism for a wide class
of real glasses, is in semiquantitative 
agreement with experimental results.  
\end{abstract}

\section{Introduction}
Raman scattering is a very powerful technique to study the vibrational
dynamics of solids, and more so in disordered solids where the
vibrational eigenvectors are not propagating plane waves and the law of
pseudo-momentum conservation does not hold. This has the consequence that
all modes, to some extent, can in principle scatter the light. From the 
experimental point of view, the Raman spectra of most glasses show two 
characteristic features at low frequency: the so-called quasi elastic 
scattering excess (QESE) in the energy range below $\approx 5 \div 
10$~cm$^{-1}$, 
and the boson peak, a relatively broad band peaked in the range 10$\div$100
cm$^{-1}$ depending on the nature of the glass \cite{jrs,terki}.
\par
The QESE manifests itself in the fact that by rising the temperature, the
scattering intensity in the indicated range 
grows more rapidly than predicted by the Bose-Einstein statistics 
(which is not the case for the  boson peak).
This indicates that the QESE originates either from two-"phonon"
processes or from anharmonic degrees of freedom consisting possibly of some
sort of generalized two-level systems \cite{wint} that are dealt with by
the so-called soft potential model \cite{gurevich}. 
At low enough temperature the QESE contribution to the scattering becomes
negligible, and what is left in the spectrum at frequency higher than a few
wavenumbers derives presumably from harmonic, acoustic-like vibrations of 
low frequency. In the following we will focus our attention on these 
vibrational modes.
\par
Starting from the work of Shuker and Gammon \cite{sg} many authors 
\cite{galeener,marbr,nemanich} with different approaches and approximations
showed that the Stokes intensity scattered at first order by harmonic
vibrations in disordered solids can be written as
\begin{equation}
   I(\omega, T) \propto \frac{1}{\omega}[n(\omega, T)+1] C(\omega) g(\omega)
\end{equation}
where $n(\omega, T)$ is the Bose-Einstein population factor, $C(\omega)$ is 
the
average light-vibration coupling coefficient of the modes having frequency
between $\omega$ and $\omega + d\omega$ \cite{nota1}, 
and $g(\omega)$ is the density of
vibrational states. 
\par
The frequency-dependence of $I(\omega, T)$, or equivalently of $C(\omega)$, 
has
been the object of much debate, and in particular the question concerns
whether they are scaling quantities or not 
\cite{boukenter,eurlett,tsujimi,noiprl,noicol,acv,duval1,momenti,terao},
both in real glasses and in model disordered systems.
The investigation of the frequency dependence of $C(\omega)$ for model
disordered solids is becoming more and more attractive also in view of the
recent demonstration, by Fontana and coworkers \cite{aldino1,aldino2}, 
that a
combination of Raman and specific heat experiments allows for the
determination of the density of vibrational states with an accuracy
comparable to neutron scattering experiments, and therefore provides 
the functional form of $C(\omega)$.
\par
In a previous paper \cite{momenti} we have studied this problem in detail
for the case of percolating networks at percolation threshold
concentration,
showing that the scaling laws, if any, are so strongly model dependent,
that it is practically impossible to extract reliable information on the 
dynamics of the system from the knowledge of $C(\omega)$.
\par
In the present work we will study percolators having a concentration 
higher than threshold: although it is not granted that these systems can 
reproduce the scattering 
properties of real disordered systems (even neglecting QESE), they are 
certainly more realistic than percolators at threshold, but share with the 
latter a basic simplicity. In this paper we devote most of our attention 
to bond percolators 
rather than to site percolators: indeed, in our opinion, these systems
could represent a reference model for covalent, newtwork forming glasses.
\par
We report the calculation of the Raman coupling coefficients 
of percolators at high mass concentration (up to 98\%), 
and show that also in these more "realistic" model solids the existence 
of scaling
behavior of the Raman coupling coefficient is highly questionable.
\par
In order to calculate the Raman coupling coefficient $C(\omega)$ it is not
enough to know the vibrational dynamics, but it is also necessary to
specify
the mechanism by which the vibrations modulate the electric polarizability
of the scattering units. In the bond-polarizability mechanism (BPOL), the
electric polarizability is localized on the bonds that link the atoms and
is
directly modulated by the change in bond lengths produced by the
vibrations.
In the dipole-induced-dipole (DID) mechanism the polarizable units are the
atoms and the vibrations, by changing the interatomic distances, modulate
the dipolar interaction between them and thus their polarizability. Very
roughly speaking, BPOL may be expected to dominate in covalent materials,
while DID has been shown to be the most important mechanism 
for rare gases and, in 
general, in the presence of van der Waals interactions. However, in real 
glasses the situation will not be that simple; for example, in a molecular 
solid like a polymer we might expect BPOL to dominate for frequencies 
corresponding to intramolecular modes and DID for frequencies corresponding
to intermolecular ones. Moreover, even for network forming glasses the two 
scattering mechanisms might coexist in the same frequency range. These 
considerations imply that one should be very cautious in establishing a 
strict correspondence between the results of simulation and experiments.
Nevertheless, due to the difficulties of simulating real glassy
samples large enough to give access to the low frequency region where the 
scaling laws are assumed to hold, we think that the simulation of 
model systems remains an important tool for the study of the dynamics
of topologically disordered solids.
\par
%%%%
\section{Results and discussion}
%%%%
The numerical method we use is the method of moments \cite{benoit,momenti};
the calculation and the expedients that ensure  good covergence with large
systems were discussed at length in ref. \cite{momenti}, together with the
different forms the equations of the moments take for BPOL and DID, and we
refer the reader to this reference for the details. 
\par
The systems we study are 3-dimensional site- and (mostly)
bond-percolators
consisting of identical masses linked by identical springs; each mass is
assigned a single vibrational degree of freedom and periodic boundary
conditions are imposed. 
Mass and/or spring disorder could easily be introduced in the calculation, 
but its effect is mainly to distinguish between acoustic-like
modes at low frequency, and optical-like ones in the high frequency range,
while here we are interested in the low frequency acoustic spectrum.
We studied different concentrations ranging from
percolation threshold to 80\% (for site percolators) and 70\% (for bond
percolators). The linear
dimension of the samples was $L=85$ \cite{nota3}. Following the procedures
described in ref. \cite{momenti} we have computed the density of
vibrational
states and the Raman coupling coefficients for the two scattering
mechanisms, $C_{DID}(\omega)$ and $C_{BP}(\omega)$. The density of states
exhibits
the usual crossover from phonon-like to fracton-like behavior already
observed in several papers. 
\par
In Fig. 1 we report the log-log plot of DID and BPOL $C(\omega)$'s for site
percolators at a mass concentration $c_M=0.8$, together with the 
density of states. 
Hereafter, the frequency $\omega$ is in units of the maximum frequency.
In the preliminary results ($c_M=0.5$) of ref. \cite{momenti} we 
observed that the phonon-fracton crossover frequency of the density 
of states is higher than that of the coupling coefficients.
A similar behavior is observed in Fig.~1; indeed: ($i$) the  
density of states follows a phonon-like behavior up to a frequency ($\omega
\approx 0.5$) higher than the crossover frequency observed in 
DID and BPOL coupling constants (i.e. $\omega \approx 0.2$), and ($ii$) 
none of
the slopes of 
$C_{DID}(\omega)$ and $C_{BP}(\omega)$ observed in the "fracton" frequency
region has direct connection with the slopes at threshold \cite{momenti}.
\par
In Fig. 2 are shown the BPOL coupling coefficients $C_{BP}(\omega)$ 
for bond percolatores at bond concentrations ranging from $c_B=0.249$ 
(percolation threshold) to $c_B=0.7$. It is worth to note that 
these two extreme {\it bond} concentrations correspond to {\it mass}
concentrations $c_M$ of 0.35 and 1 respectively (to a bond
concentration of 0.5 corresponds a mass concentration of 0.98),
so that bond percolators with $c_B$ in the range between 0.5 and 1 
could represent a model for real network forming glasses.
\par 
An interesting new feature can be observed in Fig.~2: as $c_B$ is 
increased the low frequency part of the spetrum acquires a
concentration-dependent slope, $S_L(c)$, that is appreciably {\it lower}
than the slope at 
percolation threshold ($S(c=c_T) \approx 1.6$).
This is contrary to what would be expected by analogy to the phonon-fracton
crossover observed in the density of states, which produces a {\it higher}
slope at low frequency. The crossover frequency observed in the present
case
increases with concentration. It is likely that a larger "phononic" slope
is 
present as well, but at too low a frequency to be observed in our samples.
In any case a phonon-like contribution with slope $S>2$  
starts to be observed at $c_B=0.35$, and at $c_B=0.5$ it covers
the whole low frequency part of the spectrum ($\omega < 0.1$); at this
and higher concentration, the high-frequency slope is intermediate between
those found at low frequency and at threshold.
\par
In Fig. 3 we report the $C_{DID}(\omega)$ coupling coefficients for 
bond percolators  with 
concentrations ranging from $c_B=0.249$ to $c_B=0.7$. As the concentration
increases, the low frequency "scaling" \cite{rayo} part of the
spectrum at threshold concentration is more and more covered from both
sides: from low frequency by the
"phononic" slope and from high frequency by a roundish spectrum to which no
slope can reasonably be associated. Already at a bond concentration
$c_B=0.325$, corresponding to a mass concentration $c_M \approx 0.79 $,
the threshold slope ($S(c_T) \approx 0.2)$ has completely disappeared.
\par
The main motivation of the present investigation was to check in some
detail, and on simple models, whether the claimed scaling behavior of the
low frequency coupling coefficient $C(\omega)$ would survive in a 
situation less
unrealistic than percolation threshold. 
\par
The answer is definitely negative for DID, and it may be worth to clarify
shortly why the authors of ref. \cite{terao} may have concluded that at a
concentration $c_B=0.31$ (comparable to our Fig. 3(b)) the spectrum does 
scale above the crossover.
\par
From Fig. 3(b) it is clear that the slope above 
the crossover ($S(c>c_T) \approx 0$) has nothing to do with that at low 
frequency in the threshold spectrum of Fig. 3(a) ($S(c_T) \approx 0.2$). 
The fact is that in ref. \cite{terao} it is the reduced Raman intensity, 
$$
J(\omega)=I(\omega)/n(\omega)+1 \propto C(\omega) \rho(\omega) 
\propto C(\omega) \omega^{-0.67}
$$ 
that is plotted, and not $C(\omega)$. Being $C(\omega)$ in the
frequency range of interest only slightly roundish (see Fig. 3(b)),
$J(\omega)$ will have a sligtly rounded look superimposed on a line of slope 
$ \approx -0.5$, that is exactly what is observed in ref. \cite{terao},
Fig.
2(b). The same qualitative arguments apply to their Fig. 2(a). 
The scaling of $C_{DID}(\omega)$ found in ref.~\cite{terao} is therefore an
artifact that arises from plotting $J(\omega)$, rather than
$C(\omega)$, on a shrinked vertical scale.
\par
The situation is more complex for BPOL which, as mentioned, is expected to
be the dominant mechanism for covalent solids. More than one power law
is observed in this case. The origin of the new low-frequency slope that
emerges as the concentration is increased, is not clear
to us. In any case, that slope is not very different from the one at
percolation threshold, and compares rather favourably with the values
suggested from combined Raman-neutron scattering \cite{malino} and 
Raman-thermal \cite{aldino1,aldino2} experiments, that yield 
$C(\omega) \propto \omega^{0.7 \div 1}$.
\par
Though in our opinion one should be very cautious and await calculations on
more realistic systems, it is not unconceivable that these values reflect
the slopes computed here. The fact that slopes of the order of unity are
observed
in many different glasses suggests that they are the result of very general
properties of BPOL-like scattering mechanismes in disordered media: 
therefore, after all, it is not too much of a surprise if a simple 
model like the present one has the same qualitative features. 
On the other hand the appearance of a {\it smaller} slope at very low
frequency could be an indication that, also in network forming glasses,
the coupling coefficient $C(\omega)$ reaches a constant value when $\omega$
approaches
zero. Indeed this kind of non-scaling behavior has been recently found in a
(simulated) model of fragile glass \cite{argon}.
\par
In conclusion, in the present work we have shown that a large variety of
behavior is found for $C(\omega)$ at low frequency. It is difficult to
reach a definite conclusion as to the validity of scaling laws
for BPOL, but we feel confident in asserting that extracting quantitative 
information
on the system's parameters from the measured spectra on the basis 
of scaling arguments, is potentially misleading and at present unreliable.
\par
On the other hand, the numerically computed frequency dependence of
$C_{BP}(\omega)$ is in semiquantitative agreement with experimental
findings. Such capability of simple percolating structures to
reproduce widespread properties of disordered solids was previously
pointed out by Sheng and Zhou \cite{cinesi}, who by using 
site percolators above threshold were able to reproduce
the plateau of low-temperature thermal conductivity. It might be 
interesting to check whether the other common feature
of Raman scattering in glasses, i.e. the boson peak, is a characteristic
of high-concentration percolators as well.  

\newpage

\begin{center}
{\bf FIGURE CAPTIONS}
\end{center}

\begin{description}
\item  {FIG. 1 - 
Site percolator, $L=85$, $c=0.8$, 1 realization. (a) density of states; (b)
$C_{DID}(\omega)$; (c) $C_{BP}(\omega)$. 
For graphical convenience the traces
are vertically shifted.
}
\item  {FIG. 2 -
$C_{BP}(\omega)$ for bond percolators, $L=85$, average of 
10 realizations, at
various bond concentrations. (a) $c_B=0.249$; (b) $c_B=0.325$; (c)
$c_B=0.35$;
(d) $c_B=0.4$; (e) $c_B=0.5$; (f) $c_B=0.7$.          
For graphical convenience the traces
are vertically shifted.
}
\item  {FIG. 3 - 
$C_{DID}(\omega)$ for bond percolators, $L=85$, average of 10 
realizations, at
various bond concentrations. (a) $c_B=0.249$; (b) $c_B=0.325$; (c)
$c_B=0.4$; 
(d) $c_B=0.7$.          
For graphical convenience the traces
are vertically shifted.
}
\end{description}


\begin{references}
\bibitem{nota1} 
As shown in ref.~\cite{physica}, the coupling coefficient $C_i=C(\omega_i)$ is 
a quantity associated to individual modes, and it vary by orders of 
magnitude in passing from one mode to the adjacent one.
Therefore only its average over a large enough energy range, i.e.
$C(\omega)$, ia a smooth function of $\omega$.
\bibitem{nota3} 
In a few computer runs we have used $L=120$, obtaining no
observable difference in the results apart from a lower minimum frequency.
\end{references}
\end{document}